\documentclass[aps,prl,twocolumn,superscriptaddress,reprint]{revtex4-1}

\usepackage[%
	colorlinks=true,
	urlcolor=blue,
	linkcolor=blue,
	citecolor=blue,
	breaklinks=true
]{hyperref}
\usepackage[english]{babel} % language used
\usepackage{bm}% bold math
\usepackage{graphicx} % figures
\usepackage{color}
\usepackage{siunitx} % SI units symbols

\begin{document}

\title{Experimental Determination of Momentum-Resolved Electron-Phonon Coupling}

\author{Matteo~Rossi}
\email{matteo1.rossi@polimi.it}
\affiliation{Dipartimento di Fisica, Politecnico di Milano, Piazza Leonardo da Vinci 32, I-20133 Milano, Italy}

\author{Riccardo~Arpaia}
\affiliation{Dipartimento di Fisica, Politecnico di Milano, Piazza Leonardo da Vinci 32, I-20133 Milano, Italy}
\affiliation{Department of Microtechnology and Nanoscience, Chalmers University of Technology, SE-41296 G\"oteborg, Sweden}

\author{Roberto~Fumagalli}
\affiliation{Dipartimento di Fisica, Politecnico di Milano, Piazza Leonardo da Vinci 32, I-20133 Milano, Italy}

\author{Marco~{Moretti~Sala}}
\affiliation{Dipartimento di Fisica, Politecnico di Milano, Piazza Leonardo da Vinci 32, I-20133 Milano, Italy}

\author{Davide~Betto}
\altaffiliation[Present address: ]{Max Planck Institut f\"{u}r Festk\"{o}rperforschung, Heisenbergstrasse 1, D-70569 Stuttgart, Germany}
\affiliation{ESRF -- The European Synchrotron, 71 Avenue des Martyrs, CS 40220, F-38043 Grenoble, France}

\author{Gabriella~M.~{De~Luca}}
\affiliation{Dipartimento di Fisica \textquotedblleft E. Pancini\textquotedblright, Universit\`{a} degli Studi di Napoli \textquotedblleft Federico II\textquotedblright, Complesso Monte Sant\textquoteright Angelo - Via Cinthia, I-80126 Napoli, Italy}
\affiliation{CNR-SPIN, Complesso Monte Sant\textquoteright Angelo - Via Cinthia, I-80126 Napoli, Italy}

\author{Kurt~Kummer}
\affiliation{ESRF -- The European Synchrotron, 71 Avenue des Martyrs, CS 40220, F-38043 Grenoble, France}

\author{Jeroen~{van~den~Brink}}
\affiliation{Institute for Theoretical Solid State Physics, IFW Dresden, Helmholtzstrasse 20, D-01069 Dresden, Germany}
\affiliation{Department of Physics, Technical University Dresden, D-01062 Dresden, Germany}

\author{Marco~Salluzzo}
\affiliation{CNR-SPIN, Complesso Monte Sant\textquoteright Angelo - Via Cinthia, I-80126 Napoli, Italy}

\author{Nicholas~B.~Brookes}
\affiliation{ESRF -- The European Synchrotron, 71 Avenue des Martyrs, CS 40220, F-38043 Grenoble, France}

\author{Lucio~Braicovich}
\affiliation{Dipartimento di Fisica, Politecnico di Milano, Piazza Leonardo da Vinci 32, I-20133 Milano, Italy}
\affiliation{ESRF -- The European Synchrotron, 71 Avenue des Martyrs, CS 40220, F-38043 Grenoble, France}

\author{Giacomo~Ghiringhelli}
\email{giacomo.ghiringhelli@polimi.it}
\affiliation{Dipartimento di Fisica, Politecnico di Milano, Piazza Leonardo da Vinci 32, I-20133 Milano, Italy}
\affiliation{CNR-SPIN, Dipartimento di Fisica, Politecnico di Milano, Piazza Leonardo da Vinci 32, I-20133 Milano, Italy}

\begin{abstract}
We provide a novel experimental method to quantitatively estimate the electron-phonon coupling and its momentum dependence from resonant inelastic x-ray scattering (RIXS) spectra based on the detuning of the incident photon energy away from an absorption resonance. We apply it to the cuprate parent compound NdBa$_2$Cu$_3$O$_6$ and find that the electronic coupling to the oxygen half-breathing phonon mode is strongest at the Brillouin zone boundary, where it amounts to $\sim 0.17$ eV, in agreement with previous studies. In principle, this method is applicable to any absorption resonance suitable for RIXS measurements and will help to define the contribution of lattice vibrations to the peculiar properties of quantum materials.
\end{abstract}

\maketitle

The interaction between electrons and lattice vibrations determines a variety of physical properties of condensed matter. The electron-phonon coupling (EPC) influences the temperature dependence of the resistivity in metals, can be responsible for the renormalization of the effective mass of carriers and enables optical transitions in indirect-gap semiconductors. In conventional superconductors, the EPC drives the formation of bosonic bound states of electrons (the Cooper pairs) \cite{Bardeen1957}. In unconventional superconductors such as copper oxides (cuprates), the role of the EPC is lively discussed \cite{Lanzara2001,Bohnen2003,Zhou2005,Lee2006,Reznik2006,Anderson2007,Zhang2007,Scalapino2012,Keimer2015,He2018}. Though a purely phonon-mediated mechanism cannot account for the high critical temperatures observed in doped cuprates, the electron-lattice interaction may enhance pair binding when it operates in synergy with a dominant mechanism \cite{Savrasov1996,Nazarenko1996,Shen2002,Ishihara2004,Sandvik2004,Johnston2010}. 

\begin{figure*}
	\centering
	\includegraphics[width=\textwidth]{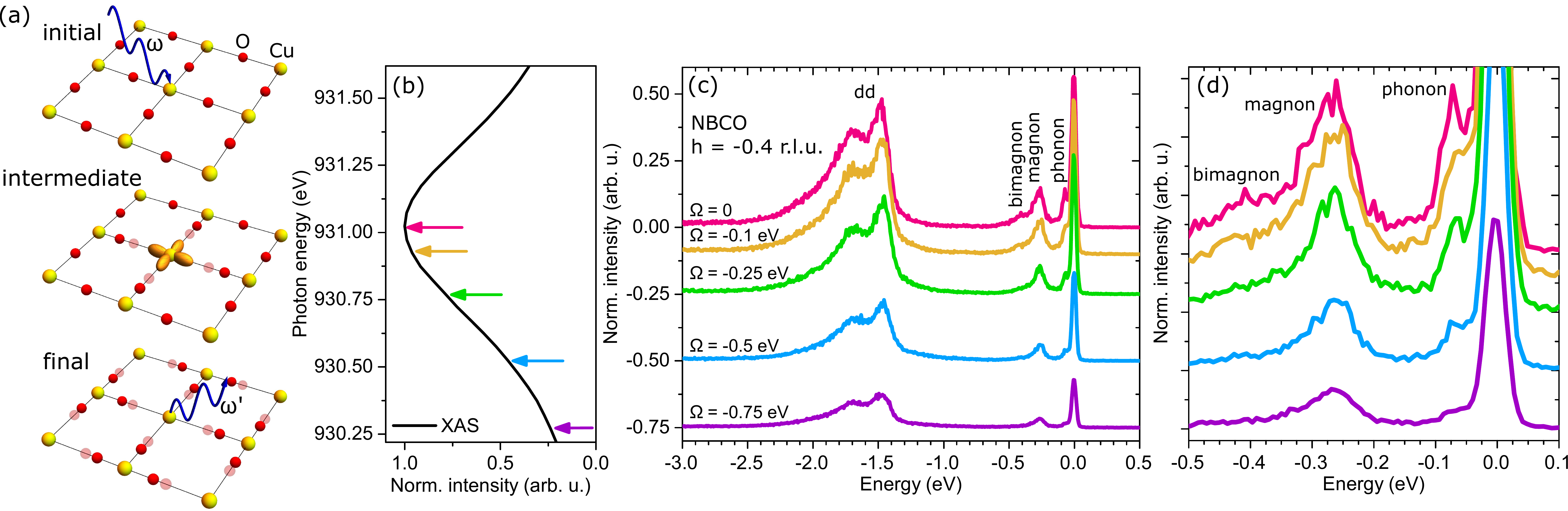}
	\caption{\label{fig1} (a) Schematics of phonon generation during the RIXS process. Phonons are created during the intermediate state when the extra electron in the $3d_{x^2-y^2}$ state locally modifies the charge density and perturbs the lattice. (b) XAS spectrum of NBCO close to the Cu $L_3$ resonance. Arrows point to the incident photon energies at which the RIXS spectra were measured. (c) RIXS spectra of NBCO measured at in-plane momentum transfer $h = -0.4$ r.l.u. as a function of the detuning energy $\Omega$. Spectra are normalized to the incident photon flux. (d) Close-up of panel (c) showing the low-energy region of the RIXS spectra.}
\end{figure*}

Experimental probes assessing the EPC include inelastic neutron and x-ray scattering (INS and IXS, respectively) \cite{Pintschovius2005,Reznik2010}, Raman spectroscopy \cite{Thomsen2006,Zhang2013}, and angle-resolved photoemission spectroscopy (ARPES) \cite{Damascelli2003,Cuk2005}. Resonant inelastic x-ray scattering (RIXS) has emerged as a complementary technique that enables to directly quantify the momentum-dependent coupling strength between a given phonon mode and the photoexcited electron \cite{Ament2011,Devereaux2016,Geondzhian2018}. The excitation of lattice vibrations during the scattering process is schematically illustrated in Fig.~\ref{fig1}(a) for the specific case of the Cu $L_3$ edge of a cuprate material. A resonant incident photon excites a $2p$ core electron into the unoccupied $3d_{x^2-y^2}$ orbital. During the intermediate state, the extra electron locally alters the charge density and repels the nearby O ions. The distorted lattice can be described by a coherent superposition of phonons. When the electron decays, the lattice is left in an excited state. The core hole is usually believed to be fully screened and therefore ineffective in the creation of phonons \cite{Ament2011,Devereaux2016}. In the limit of weak and intermediate EPC, the RIXS phonon intensity is directly proportional to the coupling between the photoexcited electron and the phonon and to the intermediate-state lifetime \cite{Ament2011}.

An estimate of the EPC on an absolute scale from RIXS spectra has been obtained from the ratio between the phonon intensity and its first overtone \cite{Ament2011,Lee2013,Moser2015,Fatale2016,Johnston2016,Meyers2018}. However, this approach is limited to measurements where phonon satellites are detected. Vibrational overtones are observed only in gaseous samples \cite{Hennies2010,Schreck2016} and systems with narrow resonances (i.e. long-lasting intermediate states, e.g. oxygen $K$ edge) and/or strong EPC \cite{Lee2013,Moser2015,Fatale2016,Johnston2016,Meyers2018}. As a matter of fact, in most RIXS spectra an unambiguous determination of phonon overtones is not possible. This Letter introduces a novel approach suitable to all absorption edges and materials for the determination of the EPC from RIXS spectra. We demonstrate that energy-detuned RIXS experiments allow to evaluate the momentum-dependent EPC in the system under study. Indeed, the EPC can be estimated by comparing the measured incidence-energy dependence of the phonon intensity with predictions based on available theoretical models. We apply the novel method to the cuprate parent compound NdBa$_2$Cu$_3$O$_6$ (NBCO). We find that the coupling strength of the electrons to the oxygen bond-stretching mode is $\sim 0.17$ eV close to the antinodal point and decreases towards the Brillouin zone center, in agreement with previous studies \cite{Johnston2010,DeFilippis2012,Devereaux2016,Farina2018}.

NBCO films with a thickness of 100 nm were deposited on single-crystalline SrTiO$_3$(100) substrates by high oxygen pressure diode sputtering, as detailed in Ref.~\cite{Salluzzo2005}. High energy resolution Cu $L_3$ edge RIXS measurements were performed at the beamline ID32 of the ESRF -- The European Synchrotron (France) using the ERIXS spectrometer \cite{Brookes2018}. The overall energy resolution was 32 meV. The incident photon polarization was set perpendicular to the scattering plane. The scattering angle was fixed at \SI{149.5}{\degree}. Measurements have been performed at 20 K. Additional experimental details are reported in the Supplemental Material \cite{SupplementalMaterial}.

Figure~\ref{fig1}(c) shows the RIXS spectra of NBCO collected at in-plane momentum transfer $\mathbf{q}_{\parallel} = (h, 0) = (-0.4, 0)$ r.l.u. as a function of the detuning energy $\Omega = \omega - \omega_\mathrm{res}$, where $\omega$ is the incident photon energy and $\omega_\mathrm{res}$ is the resonant energy [maximum of the x-ray absorption spectrum (XAS), Fig.~\ref{fig1}(b)]. The spectra are normalized to the incident photon flux and energy-dependent self-absorption effects have been estimated and corrected for \cite{SupplementalMaterial}. In RIXS spectra of undoped cuprates, sharp excitations are observed. Phonons populate the energy range down to $\sim -0.1$ eV, magnetic excitations extend until $\sim -0.6$ eV, and electronic transitions within the $3d$ manifold ($dd$ excitations) are observed between $\sim -1$ and $-3$ eV. The apparently trivial decrease of spectral intensity upon detuning is remarkably rich in information: Fig.~\ref{fig1}(c) suggests that the intensity ratio of the magnon and the $dd$ excitations is kept constant upon detuning. Instead, the close-up on the low-energy region of the RIXS spectra [Fig.~\ref{fig1}(d)] shows that the intensities of the phonon peak at $-70$ meV and of the bimagnon decrease faster than the one of the magnon. We here underline that the choice of momentum transfer allows to safely decouple magnetic and lattice excitations \cite{Pintschovius2005,Peng2017}.

\begin{figure}
	\centering
	\includegraphics[width=\columnwidth]{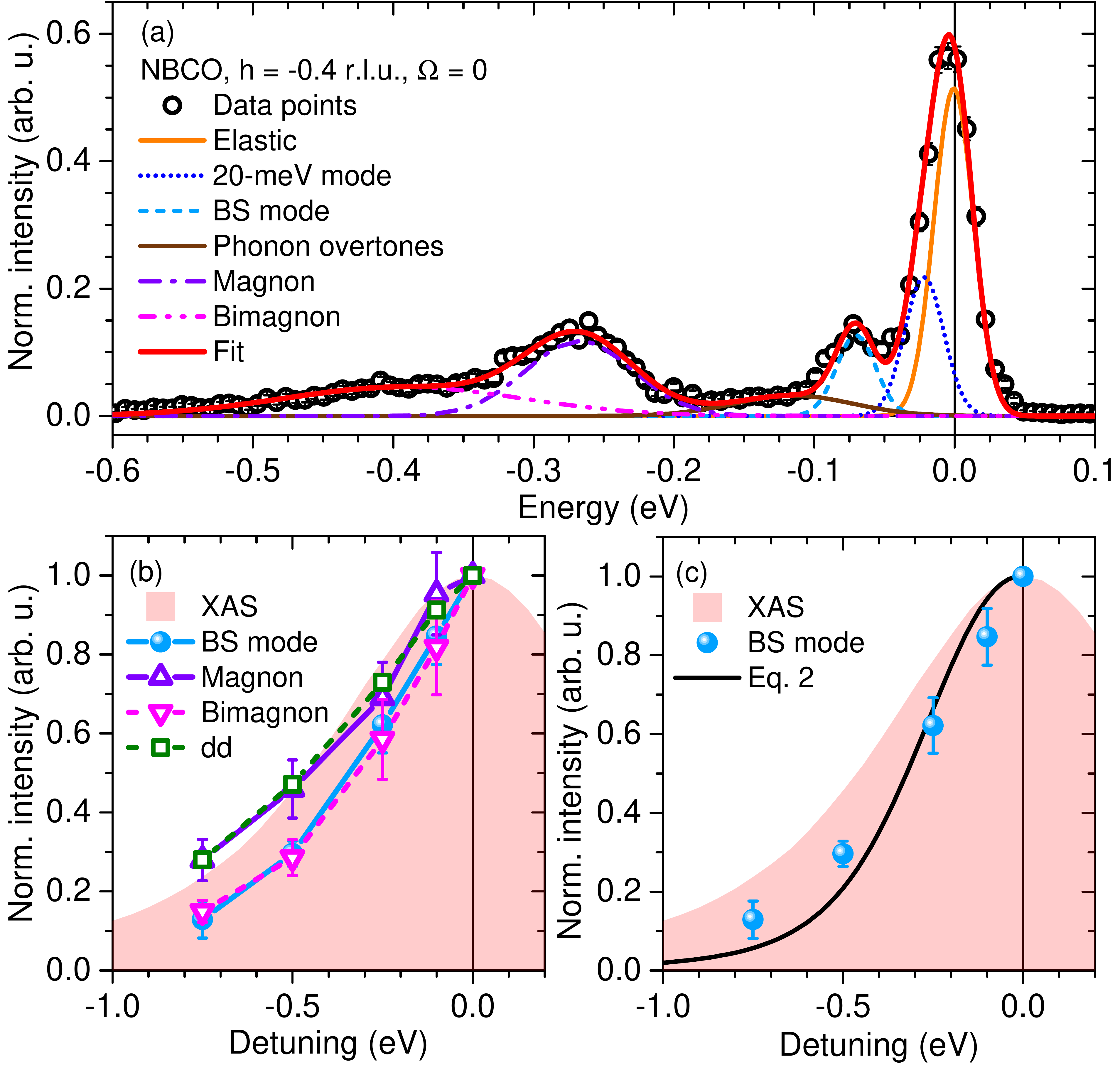}
	\caption{\label{fig2} (a) Fit (solid line) of the low-energy region of the RIXS spectrum of NBCO (circles) measured at $h = -0.4$ r.l.u. and $\Omega=0$. The single Gaussian features are also plotted. (b) Detuning dependence of the spectral weight of the excitations (symbols) compared to the XAS spectrum (shaded area). (c) Detuning dependence of the BS mode (filled circles) and expected phonon intensity dampening calculated from Eq.~\ref{eq:phonon_intensity} with parameters $\Gamma = 0.28$ eV, $\omega_0 = 0.07$ eV, and $M = 0.18$ eV (solid line).}
\end{figure}

The spectra have been fitted using Gaussian lineshapes, as shown in Fig.~\ref{fig2}(a) for the spectrum measured with $\Omega = 0$. Besides the elastic line, the fit reveals the presence of two resolution-limited excitations at $-20$ meV and $-70$ meV, and a broad distribution peaked around $-120$ meV. Following previous studies of NBCO and isostructural YBa$_2$Cu$_3$O$_{7-\delta}$ \cite{Cardona1988,Liu1988,Yoshida1990,Limonov1998,Bohnen2003,Pintschovius2005}, the excitation at $-20$ meV is given by the envelope of low-energy phonons, while the one at $-70$ meV is attributed to the Cu-O bond stretching (BS) mode (half-breathing mode). The broad distribution at 120 meV is attributed to weak unresolved phonon overtones. The single spin-flip excitation is located at $-270$ meV, while bimagnon excitations extend down to $\sim -600$ meV \cite{Peng2017}. The spectral weights of the relevant features are plotted in Fig.~\ref{fig2}(b) as a function of the detuning energy together with the XAS (shaded area), which indicates the expected intensity dampening. The weight of each feature is plotted relatively to its corresponding value at resonance. It is evident that the single magnon (upward triangle) and the \emph{dd} excitations (squares) follow the trivial intensity decrease of the XAS, while the BS mode (filled circles) and the bimagnon (downward triangles) display a faster decrease upon detuning.

In order to interpret the softening of the spectral intensities with $\Omega$, we consider the formal concept of the \emph{effective} duration $\tau$ of the scattering process and its relationship with the detuning energy. When the incident photon energy is set close to the resonance maximum ($|\Omega| \ll \Gamma$, where $\Gamma$ is the half width at half maximum of the resonance), the duration time is dictated by the intrinsic intermediate-state lifetime, according to Heisenberg's uncertainty principle: $\tau \approx 1/\Gamma$. When the incident photon energy is tuned far from the resonance ($|\Omega| \gg \Gamma$), the \emph{effective} duration of the scattering process is limited by the detuning energy: $\tau \approx 1/|\Omega|$ \cite{Williams1974}. In the intermediate regime ($|\Omega| \sim \Gamma$), one can define an \emph{effective} intermediate-state lifetime as $\tau \approx 1/\sqrt{\Omega^2 + \Gamma^2}$ \cite{Gortel1998,Gelmukhanov1999,VanDenBrink2006}. Note that the scattering time is the longest at resonance. The idea of an \emph{effective} duration has been already used, e.g., to explain the reduction of vibrational overtones in energy-detuned RIXS spectra of O$_2$: upon detuning, the system is given less time to induce multiple vibrations \cite{Hennies2010}. A detailed treatment of the \emph{effective} duration time of the scattering process falls outside of the purpose of this Letter, and the interested reader is referred to the relevant literature \cite{Williams1974,Berg1974,Skytt1996,Gelmukhanov1997,Gortel1998,Gelmukhanov1999,Gelmukhanov1999a,Ignatova2017}.

Having introduced the notion of an \emph{effective} scattering time, we classify \textquoteleft fast\textquoteright~and \textquoteleft slow\textquoteright~excitation processes on the basis of their behavior upon detuning. When the detuning energy is increased, i.e. the scattering time is decreased, \textquoteleft fast\textquoteright~processes are unaffected, and will show a trivial intensity decrease that follows the XAS profile. This is the case of magnon and $dd$ excitations, which are instantaneous processes that involve a quick redistribution of the electronic density. On the other hand, \textquoteleft slow\textquoteright~processes are penalized by a decrease of the scattering time, and will be weakened below the XAS threshold. This is what happens to phonons, which need longer excitation times since they involve the movement of the heavy nuclei. Pump-probe experiments have revealed that in cuprates the coupling of electrons to phonons occurs in an average time of $\sim 100$ fs \cite{Perfetti2007,Pashkin2010,Gadermaier2010,DalConte2012,Giannetti2016}, much longer than the core-hole lifetime ($\sim 3$ fs). It is not surprising that the phonon intensity is significantly reduced when the core-hole lifetime is effectively shortened by a factor 3 for the largest detuning. We note that the bimagnon excitation can also be classified as a slow process, contrary to the magnon excitation \cite{Bisogni2014,Tohyama2018}. In this Letter, we will focus our attention on the BS phonon mode.

In order to quantitatively estimate the EPC, we refer to the RIXS phonon model developed by Ament \emph{et al.} \cite{Ament2011,Ament2010}. For convenience, we report here the relevant equations. The Hamiltonian $\mathcal{H}$ that couples a single electronic state to a single Einstein oscillator of energy $\omega_0$ is \cite{Ament2010,Ament2011}:
\begin{equation}
\mathcal{H} = \sum_{i}\omega_0 b_i^\dagger b_i + M c_i^\dagger c_i (b_i^\dagger + b_i),
\label{eq:Hamiltonian}
\end{equation}
where the sum runs over all sites $i$; $b^\dagger$ ($b$) and $c^\dagger$ ($c$) are the creation (annihilation) operators for phonons and electrons, respectively; and $M$ parametrizes the EPC strength. The assumption of Einstein phonon modes is a crude approximation, but it allows to derive a simple analytical expression for the RIXS phonon intensity \cite{Ament2010,Ament2011}:
\begin{equation}
I_{ph} \propto \frac{e^{-2g}}{g}\left| \sum_{n=0}^{\infty}\frac{g^n (n - g)}{n!\left[\Omega + i\Gamma + (g - n) \omega_0 \right]}\right|^2,
\label{eq:phonon_intensity}
\end{equation}
where $g = (M/\omega_0)^2$ is the dimensionless EPC strength. By constraining $\Gamma$ to 0.28 eV for Cu $L_3$ edge \cite{Krause1979} and $\omega_0$ to 0.07 eV as determined from the RIXS data, the only free parameter is the coupling strength $M$. By mapping the decrease of the phonon intensity upon detuning, the EPC is obtained. A least-squares fit of the data gives the optimal value of $M = (0.18 \pm 0.03)$ eV. Figure~\ref{fig2}(c) displays the detuning dependence of the phonon spectral weight (filled circles) and the intensity drop calculated from Eq.~\ref{eq:phonon_intensity} by setting $M$ to the optimal value (solid line).

\begin{figure}
	\centering
	\includegraphics[width=\columnwidth]{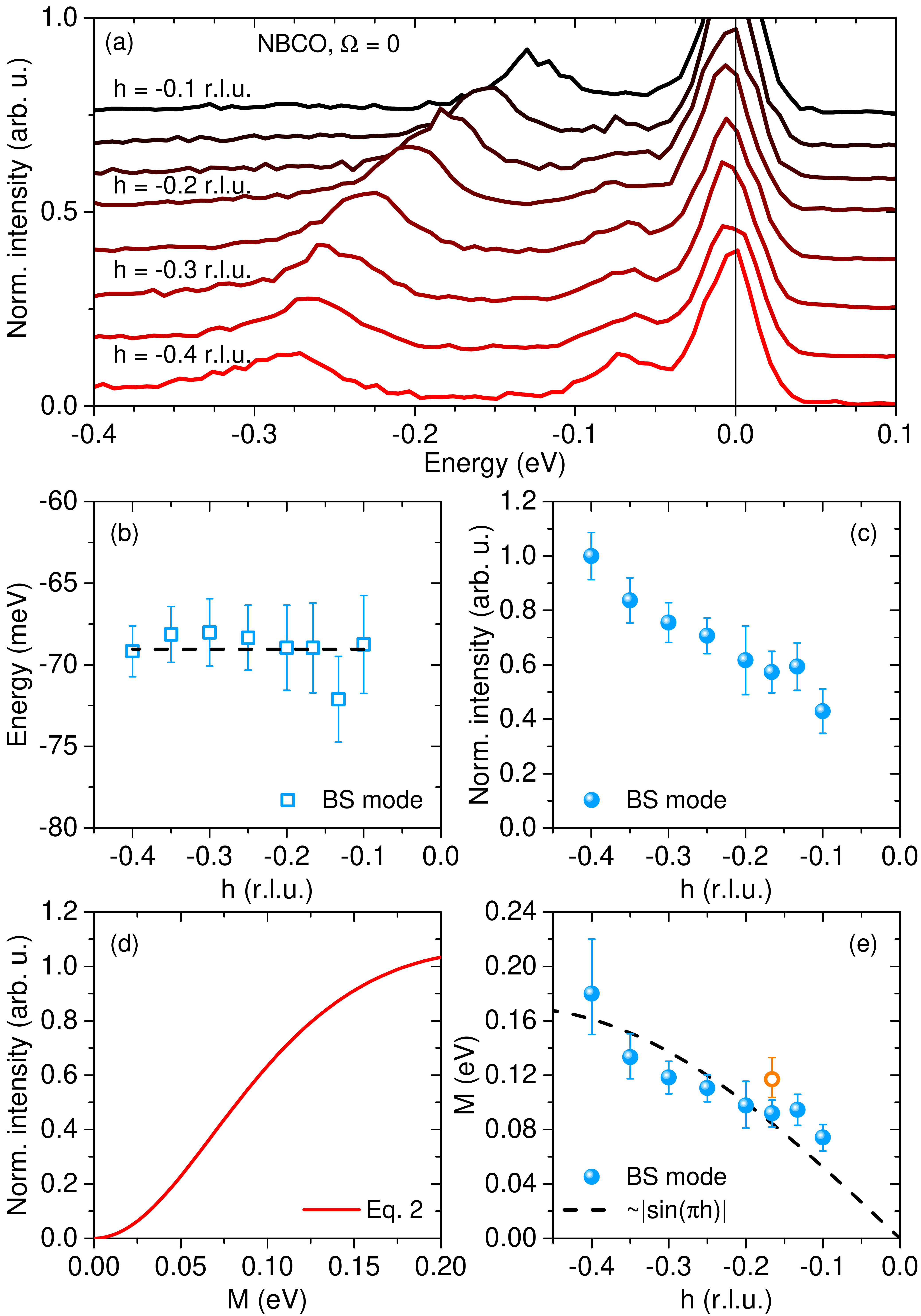}
	\caption{\label{fig3} (a) RIXS spectra of NBCO measured at resonance ($\Omega = 0$) as a function of in-plane momentum transfer $h$. Dispersion relation (b) and spectral weight (c) of the BS mode as obtained from the fit of the data. (d) Theoretical phonon intensity dependence on the coupling strength computed with $\Omega = 0$, $\Gamma = 0.28$ eV, and $\omega_0 = 0.07$ eV. The intensity is set to 1 at M = 0.18 eV. (e) Momentum dependence of the EPC (filled circles) compared to the theoretical scaling law of the coupling to the BS mode according to Ref.~\onlinecite{Song1995}. The open circle is the EPC estimated from energy-detuned RIXS measurements at $h = -0.166$ r.l.u.}
\end{figure}

Once the EPC has been calibrated at a specific momentum transfer, it is possible to obtain its momentum dependence. The procedure is illustrated in Fig.~\ref{fig3}. Panel (a) displays a series of RIXS spectra of NBCO measured with $\Omega = 0$ at different momentum transfers along the $(1, 0)$ direction. The fit of the data (not shown) allows for the extrapolation of the dispersion [Fig.~\ref{fig3}(b)] and the spectral weight [Fig.~\ref{fig3}(c)] of the BS mode as a function of $h$. The spectral weight at $h = -0.4$ r.l.u. has been set to unity. By comparing the relative change of the measured phonon intensity to the relative intensity change as a function of $M$ computed with Eq.~\ref{eq:phonon_intensity} [Fig.~\ref{fig3}(d)], the momentum dependence of the EPC is extracted. The result is reported in Fig.~\ref{fig3}(e) (filled circles). We neglect the role of the RIXS matrix elements since the dominant contribution coming from the parallel polarization channel is constant along the $(1, 0)$ direction \cite{MorettiSala2011,Fumagalli2019}.

The internal coherence of our approach is tested by evaluating the EPC strength from energy-detuned RIXS spectra measured at a different momentum, namely $h = -0.166$ r.l.u. \cite{SupplementalMaterial}. We obtain $M \approx 0.12$ eV [open circle in Fig.~\ref{fig3}(e)], in fairly good agreement with the estimated scaling law. The discrepancy (20\%) may be due to an inaccurate evaluation of the EPC at $h = -0.166$ r.l.u. resulting from a RIXS phonon signal already weak at $\Omega = 0$ and even weaker upon detuning.

The strength \cite{DeFilippis2012,Farina2018} and momentum dependence \cite{Song1995,Bulut1996,Johnston2010} of the EPC is consistent with previous studies. The scaling law of the EPC strength of the BS mode can be derived from symmetry grounds and is found to be $\propto |\sin(\pi h)|$ \cite{Song1995,Bulut1996,Johnston2010}. This behavior is reported in Fig.~\ref{fig3}(e) (dashed line), where the overall scaling factor of 0.17 eV is obtained through a least-squares fit. The good agreement between our results and previous works makes us confident on the validity of our approach. The relatively large value of the coupling strength of the BS mode ($M/\omega_0 \sim 2.5$) renews the interest in the role of the EPC in cuprates. Though it is established that coupling to phonon modes alone cannot explain the high critical temperatures of superconducting cuprates, several studies point out that lattice vibrations may cooperate with a dominant pairing mechanism and enhance superconductivity \cite{Savrasov1996,Nazarenko1996,Shen2002,Ishihara2004,Sandvik2004,Johnston2010}. Moreover, some controversies are found related to the role of the half-breathing phonon branch, which may give rise to an attractive \cite{Shen2002,Ishihara2004} or repulsive \cite{Bulut1996,Sandvik2004} interaction. 

Our pioneering work paves the way to future studies of phonons in cuprates and related quantum materials, that will provide complementary information with respect to nonresonant inelastic scattering and ARPES. Indeed, INS, IXS and Raman spectroscopy measure spectral functions of a single phonon mode influenced by scattering with all electrons in the system, while ARPES measures the coupling of a single electronic state to all phonon modes \cite{Reznik2010}. In RIXS, the unobserved intermediate state prevents a complete characterization of the specific electronic state that couples to the phonon mode in terms of momentum and energy relative to the Fermi surface. Nevertheless, the direct link between the RIXS phonon intensity and the EPC allows to recover unmatched quantitative information on the coupling strength from RIXS spectra. 

To summarize, we demonstrated that energy-detuned RIXS experiments offer a novel way to determine the EPC in a quantitative manner. Our approach has been used to find the coupling strength between the photoexcited electron and the Cu-O BS mode in the cuprate parent compound NBCO. From the relative change of the phonon intensity as a function of momentum transfer, the momentum dependence of the EPC has been determined. Since this method can be straightforwardly extended to other phonon branches and compounds, energy-detuned RIXS measurements may represent a powerful tool to extract unique information on the influence of the lattice excitations on the properties of several classes of quantum materials. RIXS has played a major role in the study of the electronic and magnetic dynamics of complex systems. In view of the design and construction of new instruments with unprecedented resolving power, this technique is expected to play a pivotal role in the study of electron-lattice interaction in the near future. 

\section*{Acknowledgments} 
The authors are grateful to K.~Gilmore, M.~Krisch and G.~Monaco for helpful discussions. RIXS data were collected at the beamline ID32 of the ESRF using the ERIXS spectrometer jointly designed by the ESRF and the Politecnico di Milano. This work was supported by the ERC-P-ReXS project (2016-0790) of the Fondazione CARIPLO and Regione Lombardia, in Italy. R.~A. is supported by the Swedish Research Council (VR) under the project \textquotedblleft Evolution of nanoscale charge order in superconducting YBCO nanostructures\textquotedblright.

\bibliography{biblio}

\end{document}